\documentclass[aps,prl,twocolumn,showpacs,superscriptaddress,groupedaddress]{revtex4} 
\pdfoutput=1
\usepackage[version=3]{mhchem} 
\usepackage[dvips]{epsfig}
\usepackage{graphicx}  
\usepackage{dcolumn}   
\usepackage{bm}        
\usepackage{amssymb}   
\usepackage{color}
\usepackage{float}

\begin{document}
\author{Aaron C. Hryciw}
\affiliation{National Institute for Nanotechnology, Edmonton, AB, T6G 2M9, Canada}
\author{Marcelo Wu}
\author{Behzad Khanaliloo}
\author{Paul E. Barclay}
\email{pbarclay@ucalgary.ca}
\affiliation{National Institute for Nanotechnology, Edmonton, AB, T6G 2M9, Canada}
\affiliation{Institute for Quantum Science and Technology, University of Calgary, Calgary, AB, T2N 1N4, Canada}
\title{Near-field tuning of optomechanical coupling in a split-beam nanocavity}

\newcommand{\dif}{\text{d}}


\begin{abstract}
Tunable evanescent coupling is used to modify the optomechanical interactions within a split-beam photonic crystal nanocavity.  An optical fiber taper probe is used to renormalize the optical nanocavity field and introduce a dissipative optomechanical coupling channel, reconfiguring and enhancing coupling between the optical and mechanical resonances of the device.  Positioning of the fiber taper allows preferential coupling to specific mechanical modes and provides a mechanism for tuning the optomechanical interaction between dissipative and dispersive coupling regimes.
\end{abstract}

\maketitle

\noindent

Photonic crystal nanocavity optomechanical devices  \cite{ref:eichenfield2009apn, ref:chan2009omd, ref:eichenfield2009oc, ref:deotare2011aor} confine light within wavelength-scale volumes, where it can interact with co-localized nanomechanical resonances. Among many recent demonstrations, these devices have been used for sensing \cite{ref:anetsberger2009nfc, ref:srinivasan2011oti, ref:krause2012ahm, ref:sun2012fdc, ref:kim2013nto,ref:wu2014ddo},  integrated photonics  \cite{ref:deotare2011aor, ref:bagheri2011dmn}, and fundamental studies of quantum nanomechanical structures  \cite{ref:chan2011lcn}.  The optomechanical coupling strength characteristic of these devices is geometry dependent; it may vanish if the spatial symmetry of the optical and mechanical resonances of the device differ.  Here we demonstrate that evanescent optical coupling between an optical fiber taper and a nanocavity can be used to reconfigure the optomechanical properties of a device. By adjusting the fiber--nanocavity geometry,  we renormalize the nanocavity optical mode, modifying both its sensitivity to individual mechanical modes of the device and the balance of dissipative and dispersive optomechanical coupling processes  \cite{ref:elste2009qni,ref:weiss2013qll}.  We show that these effects allow readout of mechanical resonances with no intrinsic optomechanical coupling---including out-of-plane cantilever modes used in atomic force microscopy and magnetometry applications  \cite{ref:bleszynskijayich2009pcn}---enabling a wider variety of mechanical resonances to be utilized for optomechanical sensing applications.  Furthermore, these effects are shown to allow spatially selective measurement of mechanical resonances, providing information describing the spatial localization of individual mechanical resonances.

\begin{figure}[htb]
\begin{center}
\epsfig{figure=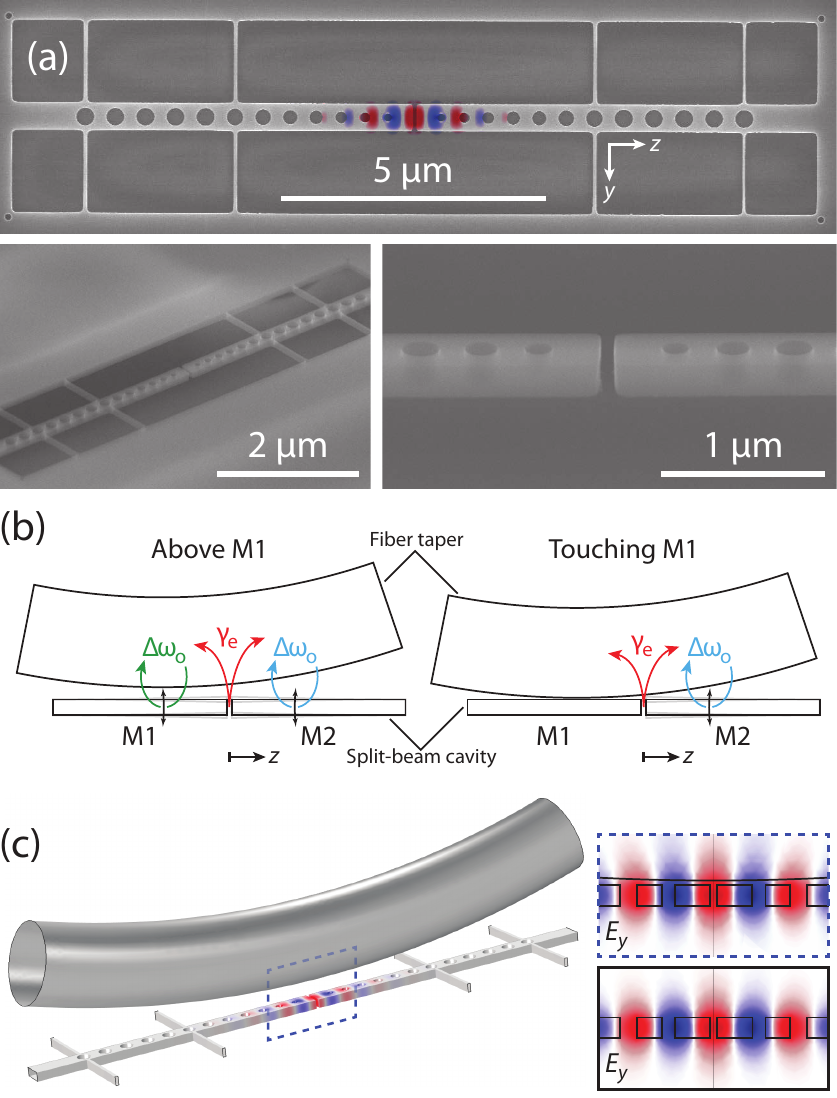, width=0.9\linewidth}
 \caption{(a) Scanning-electron micrographs of a split-beam photonic crystal nanocavity.  The nanocavity optical mode ($E_y$) is superimposed on the device in the upper image.  (b) Schematic of experimental geometry when fiber is hovering above (left) and touching (right) one of the mirrors.  (c) Renormalization of the optical mode by the optical fiber taper.  The field plots show $E_y$ in the center of the cavity with (upper) and without (lower) the fiber taper.}
\label{fig:schematic}
\end{center}
\end{figure}

In cavity optomechanical systems, mechanical excitations perturb the local dielectric environment, modifying the dynamical properties of the optical resonances \cite{ref:aspelmeyer2013co}. In nanophotonic cavity optomechanics, this modification is typically dispersive, quantified by the dispersive optomechanical coupling coefficient, $g_\text{om} = \dif\omega_\text{o}/\dif x$, which describes the change in the optical resonance frequency, $\omega_\text{o}$, for a displacement, $x$, of the mechanical mode \cite{ref:eichenfield2009oc}. For mechanical modes involving vertical out-of-plane motion, such as cantilever resonances,  $g_\text{om} = 0$ due to the even vertical symmetry of the nanocavity optical field intensity, $|E|^2$, and the odd vertical symmetry of the mechanical-resonance-induced dielectric perturbation, $\Delta\epsilon(\mathbf{r};x)$.   While fabrication imperfections or the presence of a substrate \cite{ref:li2008hof} can break this symmetry, the resulting $g_\text{om}$ is typically small. Here we demonstrate that an optical fiber taper waveguide placed in the near field of the nanocavity can introduce both large dispersive and dissipative optomechanical coupling, whose magnitude can be tuned by adjusting the fiber position.  The  dispersive coupling results from the fiber distorting the vertical profile of the nanocavity mode. Dissipative coupling \cite{ref:elste2009qni, ref:li2009rco, ref:wu2014ddo} is  due to mechanical motion modulating the fiber--nanocavity distance, and is described by $g_\text{e} = \dif\gamma_\text{e}/\dif x$, where $\gamma_\text{e}$ is the nanocavity  optical mode energy decay rate into the waveguide.  We use these effects to realize optomechanical coupling to a cantilever directly integrated within a photonic crystal nanocavity. We demonstrate that the tunable nature of this effect allows the dominant optomechanical transduction mechanism to be switched between dispersive to dissipative, and enables optomechanical coupling to be configured to enhance readout of specific mechanical modes of the device.

\begin{figure}[tb]
\begin{center}
\epsfig{figure=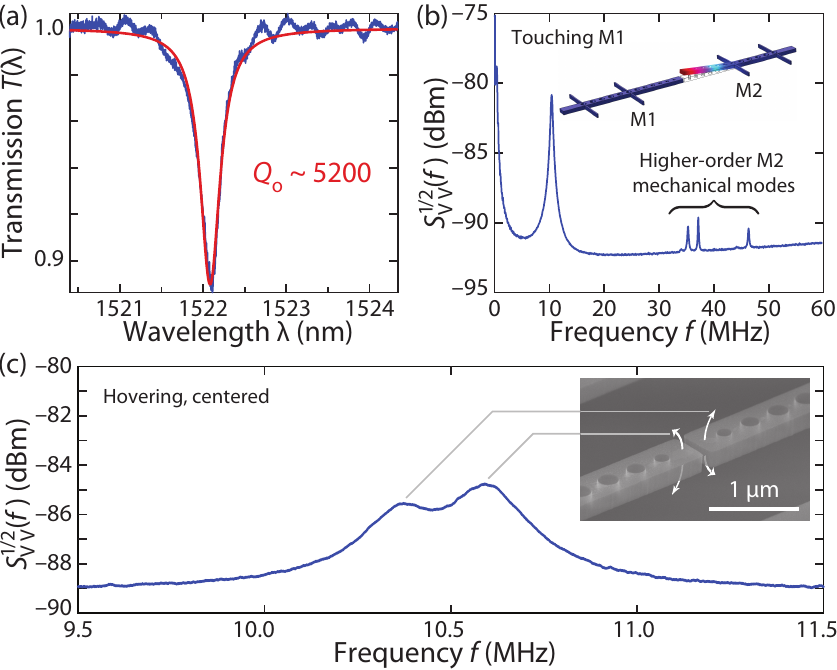, width=1\linewidth}
 \caption{(a) Fiber taper transmission under weak coupling. (b) Mechanical mode spectrum when the fiber is in contact with M1.  The displacement profile of the fundamental cantilever mode of M2 is shown next to the corresponding peak (amplitude greatly exaggerated). (c) Mechanical mode spectrum  with the fiber hovering above the cavity center ($z_\text{f}\sim 0$); fabrication imperfections impart a $\sim$200 kHz splitting between these modes.}
\label{fig:spectrum}
\end{center}
\end{figure}

Tunable optomechanical coupling is studied here using a split-beam photonic crystal nanocavity \cite{ref:hryciw2013ods,  ref:wu2014ddo}, an example of which is shown in the scanning electron microscopy (SEM) image in Fig.\ \ref{fig:schematic}(a).  Devices were fabricated from silicon using electron-beam lithography and reactive-ion etching to pattern the Si device layer of a silicon-on-insulator wafer, followed by selectively removing the underlying SiO$_2$ layer using HF wet etching.  To design a split-beam cavity with a small mode volume and relatively robust performance against fabrication imperfections, we started with the grating-defect resonator design paradigm of Liu and Yariv \cite{ref:liu2012dcr}, comprising end-to-end tapered Bragg gratings around a central defect which imparts a quarter-wave phase shift to the grating coupling coefficient.  A split-beam cavity may then be formed by incorporating a defect with a physical gap ($\sim$80 nm) at cavity center; the total defect length is chosen to minimize radiative loss, as determined from finite-difference time-domain (FDTD) simulations \cite{ref:oskooi2010mff}.  This design supports a mode with a high optical quality factor ($Q_\text{o} \sim 10^4$) at a wavelength $\lambda_\text{o} \sim 1600$ nm, whose field profile overlaps strongly with the central gap region, as shown in Fig.\ \ref{fig:schematic}(a).  We note that an alternative design approach using elliptical holes and yielding resonant frequencies near the grating band edge is capable of much lower loss ($Q_\text{o}>10^6$), but requires stringent control on hole dimensions \cite{ref:hryciw2013ods}.  Split-beam nanocavities support a variety of mechanical resonances, whose properties depend on how the mirrors are anchored to the surrounding chip.  In the device studied here, each mirror is anchored in five locations, supporting a fundamental cantilever mode with a frequency $f_\text{m} \sim 10$ MHz and displacement profile shown in the inset to Fig.\ \ref{fig:spectrum}(b).  In contrast, a single mirror support may be used to enable efficient transduction of in-plane torsional modes for magnetometry applications, as in Ref.\ \cite{ref:wu2014ddo}.

The optomechanical properties of these devices were probed using a dimpled optical fiber taper waveguide evanescently coupled to the nanocavity near field  \cite{ref:michael2007oft}.  The dimpled taper was positioned using high-resolution (50 nm) motorized stepper stages, allowing it to be hovered above the nanocavity or placed in contact with either of the nanocavity mirrors (Fig.\ \ref{fig:schematic}(b)).  The taper transmission, $T(\lambda)$, was measured by coupling a swept-wavelength laser to the fiber input and measuring the fiber output with a photodetector (Newport 1621). Optomechanical coupling was studied by measuring the spectral density of the photodetected taper transmission, $S_\text{VV}(f,\lambda)$, using a high-speed photoreceiver (Newport 1811) with a real-time spectrum analyzer (Tektronix RSA5106A).

Figure \ref{fig:spectrum}(a) shows $T(\lambda)$ when the fiber taper is hovering $\sim$500 nm above the nanocavity.  The sharp dip in transmission at $\lambda_\text{o} \sim 1522$ nm results from evanescent coupling between the fiber taper and the optical mode of the nanocavity. From the linewidth, $\delta\lambda$, and minimum transmission, $T_\text{d}$, of this resonance, the loaded and unloaded quality factors of the device are measured to be $Q_\text{o} \sim 5200$ and $Q_\text{i+P} \sim 5500$, respectively.  Figure \ref{fig:spectrum}(b) shows $S^{1/2}_\text{VV}(f)$ of the measured fiber taper transmission signal when the input laser is red detuned at $\lambda  - \lambda_o \sim \delta\lambda/2$ and the fiber taper is in contact with one of the nanocavity mirrors, labeled M1.  Several sharp resonances are visible, each corresponding to optomechanical transduction of the thermal motion from mechanical resonances of the mirror not in contact with the fiber, labeled M2.

The large peak in  $S^{1/2}_\text{VV}(f)$ at $f_\textrm{m} \sim 10.5$ MHz shown in Fig.\ \ref{fig:spectrum}(b) is from thermal motion and subsequent optomechanical coupling from the cantilever (C) mode of M2.  When the fiber taper is hovered over the center of the cavity such that it is not in contact with M1 or M2, optomechanical coupling is still present, as shown in Fig.\ \ref{fig:spectrum}(c). In the hovering configuration, the peak in $S^{1/2}_\text{VV}(f)$  has a lower amplitude and a double-peaked structure. Each of these local maxima can be ascribed to optomechanical coupling between the nanocavity and the C mode of M1 and M2, whose mechanical frequencies, $f_1 = 10.4$ MHz and $f_2 = 10.6$ MHz, respectively, differ due to fabrication variations. In an ideal structure, small displacements of the C mode would not be transduced by the optical field, as the average refractive index sampled by the optical field varies quadratically as one of the mirrors is displaced vertically.  This is a result of the opposite vertical symmetry of the mechanical mode displacement field (odd) and the optical mode energy density (even).  Non-zero optomechanical coupling can be introduced by breaking the vertical symmetry of the structure. Although this can be achieved through intrinsic fabrication imperfections, the fiber taper waveguide provides an effective method for symmetry breaking via position-dependent dissipation into the waveguide and a renormalization of the nanocavity field.  

\begin{figure}[tb]
\begin{center}
\epsfig{figure=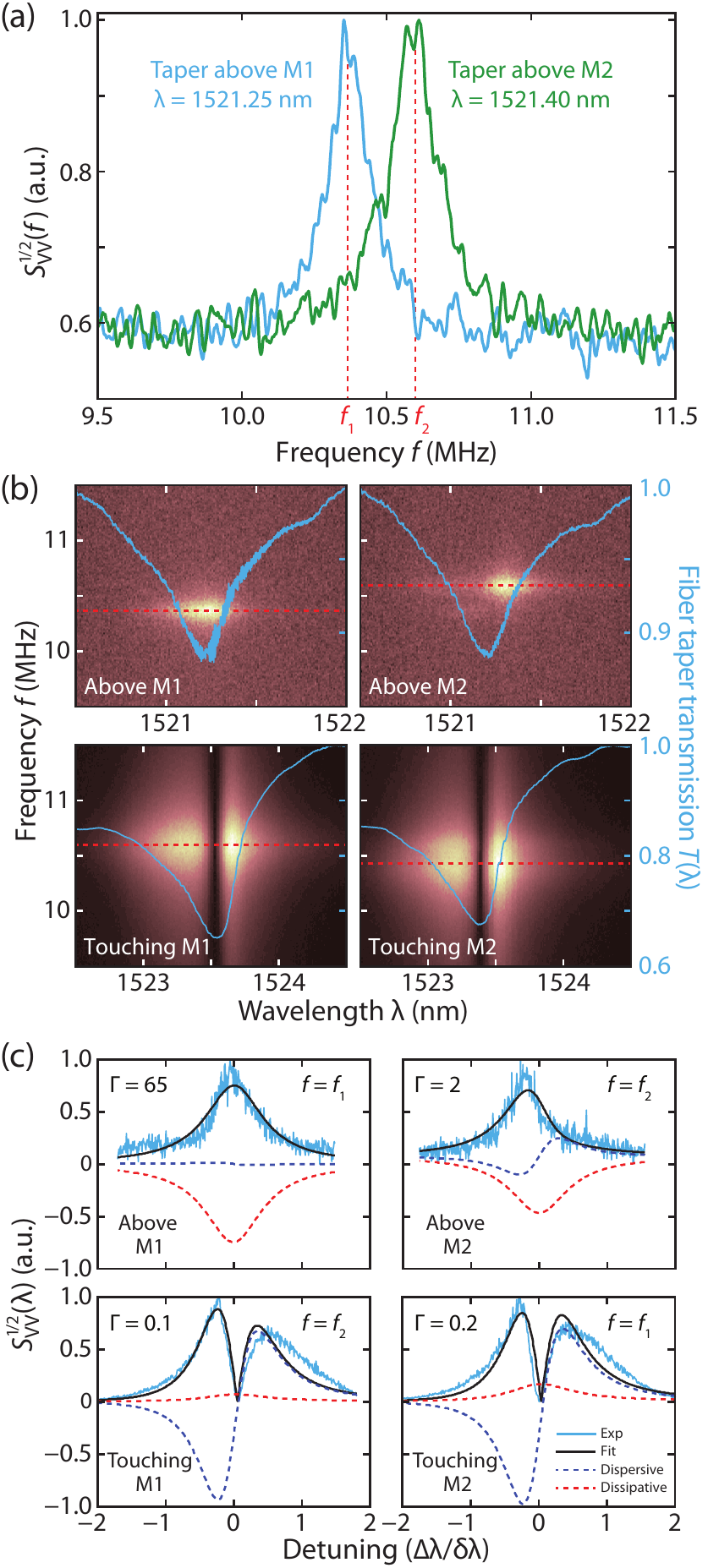, width=0.80\linewidth}
 \caption{(a) Mechanical spectrum $S^{1/2}_\text{VV}(f)$ of cantilever modes with fiber hovering $\sim$250 nm above M1 (blue) and M2 (green).  In each case, the wavelength was tuned to maximize the mechanical resonance. (b) $S^{1/2}_\text{VV}(f,\lambda)$ for the fiber hovering above/touching M1/M2, with the maxima with respect to $f$ marked by the red dotted lines.  The DC fiber transmission for each configuration is shown in blue.  (c)  $S^{1/2}_\text{VV}$ vs.\ detuning, $\Delta\lambda=\lambda-\lambda_o$, corresponding to the dotted-line slices in (b), with fits using the model in \cite{ref:wu2014ddo};  $\Gamma$ denotes the ratio of dissipative to dispersive signal amplitudes, from Eq.\ \eqref{GammaEq}.}
\label{fig:optomechanics}
\end{center}
\end{figure}

To gain insight into the optomechanical coupling processes responsible for the observed behavior, the nanocavity optomechanical response was measured as a function of axial fiber position $z_\text{f}$. When the fiber dimple is offset from the center of the nanocavity such that it is hovering above M1 ($z_\text{f}\approx -2~\mu$m), a single peak at $f_1$ was observed, as shown in Fig.\ \ref{fig:optomechanics}(a).  Similarly, when the fiber hovers above M2 ($z_\text{f}\approx 2~\mu$m), a single peak at $f_2$ appears.  These measurements indicate that the observed optomechanical coupling is not intrinsic to the optical and mechanical modes of the nanocavity alone:  the fiber position influences the optomechanical coupling processes.  

The mechanism responsible for these observations can be revealed from the $\lambda$ dependence of $S^{1/2}_\text{VV}(f,\lambda)$.  Figure \ref{fig:optomechanics}(b) shows $S^{1/2}_\text{VV}(f,\lambda)$ for four different fiber taper configurations:  hovering above or touching M1 or M2.  In all of the measurements, the maxima in $S^{1/2}_\text{VV}(f)$, marked by the red dotted lines, were observed at either $f_1$ or $f_2$.  In Fig.\ \ref{fig:optomechanics}(c), fitting $S^{1/2}_\text{VV}(f_{1,2},\lambda)$ following the procedure in \cite{ref:wu2014ddo} yields the dispersive and dissipative contributions to the total optomechanical signal.

For the hovering measurements, as $\lambda$ is tuned toward the optical resonance, optomechanical coupling is observed at $f_1$ and $f_2$ when the fiber is positioned above M1 and M2, respectively, as in the fixed-$\lambda$ measurements in Fig.\ \ref{fig:optomechanics}(a).  The maxima in $S^{1/2}_\text{VV}(f_{1,2},\lambda)$ are near $\lambda \sim \lambda_\text{o}$, indicating that the optomechanical transduction mechanism is dominantly dissipative \cite{ref:tarabrin2013adb}.  This is in contrast to the more commonly encountered dispersive coupling scenario observed in many nanophotonic cavity optomechanical systems, for which the optomechanical actuation vanishes when $\lambda = \lambda_\text{o}$. This dissipative optomechanical coupling is a result of the fiber--nanocavity coupling rate being modulated by the oscillating vertical displacement of each mirror's C mode.

\begin{figure}[h] 
\begin{center}
\epsfig{figure=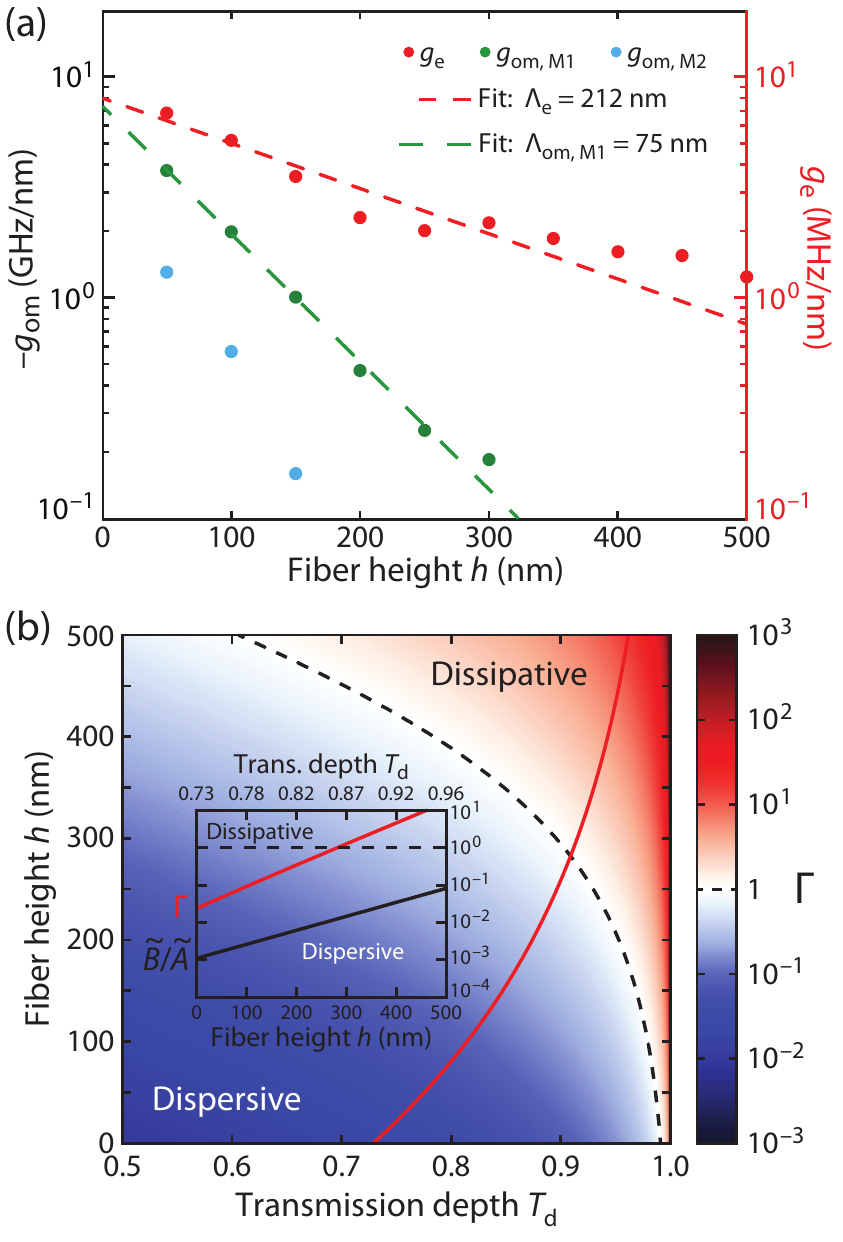, width=0.9\linewidth}
 \caption{(a) Simulated dispersive and dissipative coupling coefficients for the fiber dimple hovering above M1 ($z_\text{f} = -2~\mu$m). (b) Simulated $\Gamma(h,T_\text{d})$, the ratio of dissipative to dispersive contributions to $S^{1/2}_\text{VV}(\lambda)$.  The approximate $(T_\text{d}(h),h)$ trajectory  traversed  by the fiber taper during the experiment is shown in red.  The inset compares $\Gamma$ (red) to the ratio of dissipative to dispersive normalized coupling coefficients $\tilde{B}/\tilde{A}$ (black) along this trajectory \cite{ref:elste2009qni,ref:weiss2013qll}.}
\label{fig:simulations}
\end{center}
\end{figure}

When the dimpled fiber taper is in contact with mirror M1 (M2), the resonance at $f_2$ ($f_1$) is observed to dominate $S^{1/2}_\text{VV}(f,\lambda)$.  In this geometry, the fiber taper interacts with the optical near field of the non-contacted mirror, with a separation determined by the specific shape of the fiber taper dimple, and damps the mechanical motion of the contacted mirror.  In contrast to the hovering fiber taper geometry, the $\lambda$ dependence of  $S^{1/2}_\text{VV}(f,\lambda)$  in this configuration, shown in the bottom of Fig.\ \ref{fig:optomechanics}(c), is observed to follow $\dif T(\lambda)/\dif\lambda$:  it vanishes near resonance, indicating that dispersive coupling is the dominant optomechanical transduction mechanism.  For small gaps between the fiber taper and the free nanocavity mirror, the presence of the fiber taper creates a vertically asymmetric dielectric environment (Fig.\ \ref{fig:schematic}(c)), renormalizing the nanocavity optical mode profile and creating non-zero dispersive optomechanical coupling.  Note that this renormalization manifests in a static red-shift of $\lambda_\text{o}$ by $\sim 2$ nm, as seen in Fig.\ \ref{fig:optomechanics}(b).

The transition between predominantly dissipative and dispersive transduction can be interpreted through finite element analysis (FEA) of the optomechanical coupling as a function of vertical fiber taper position, for $z_\text{f} = -2~\mu$m (above M1).  Figure \ref{fig:simulations}(a) compares the predicted $g_\text{e}$ and $g_\text{om}$ as a function of taper height, $h$, above the nanocavity.   These simulations predict that for M1, the dispersive $g_\text{om}(h)$ decays exponentially with $h$ from a maximum absolute value of $\sim 3$ GHz/nm, following $g_\text{om} \sim g_\text{om}^\circ e^{-h/\Lambda_\text{om}}$ with $\Lambda_\text{om} \sim  75$ nm.  The dissipative $g_\text{e}$ decreases from a maximum value $\sim 10$ MHz/nm with decay length $\Lambda_\text{e} \sim 212$  nm. As the dispersive $g_\text{om}$ decays more quickly than $g_\text{e}$, for small $h$, contributions to the optomechanical coupling from $g_\text{om}$ dominate the optomechanical signal, while for large $h$, contributions from $g_\text{e}$ may dominate.  

The differing $h$ dependence of $g_\text{om}$ and $g_\text{e}$ can be understood intuitively from perturbation theory (see Supplemental Material, \S3).  In brief, the fiber  perturbs the  effective index of the nanocavity, causing a red-shift to the cavity resonant frequency $\omega_o$ which scales with the overlap of the nanocavity evanescent field intensity and the  fiber dielectric \cite{ref:johnson2002ptm}; $g_\text{om}$ consequently shares the exponential decay length of the nanocavity evanescent field intensity.  In contrast, $\gamma_\text{e}$ follows from mode coupling between the fiber and nanocavity field amplitudes \cite{ref:manolatou1999cma}, and therefore contains interference effects inherent to phase matching in addition to depending on the overlap of the evanescent tail of the fiber mode with the cavity.  As such, the exponential decay of $g_\text{e}$ depends critically on the effective coupling length between the dimpled fiber and the cavity, and is generally slow compared to the decay of $g_\text{om}$.

The effect of this behavior on the overall character of the optomechanical transduction is analyzed theoretically in Fig.\ \ref{fig:simulations}(b).  In the sideband-unresolved regime applicable here ($\omega_\text{m} \ll \gamma_\text{t}$), the maximum dispersive and dissipative contributions to the change in optical transmission are $\frac{\dif T}{\dif\omega_\text{o}}\big|_\text{max} = (1-T_\text{d})Q_\text{o}/\omega_\text{o}$ and $\frac{\dif T}{\dif\gamma_\text{e}}\big|_\text{max} = -8T_\text{d}Q_\text{o}/\omega_\text{o}$, respectively (see Supplemental Material, \S1 and Ref.\ \cite{ref:wu2014ddo}). The relative balance of  observed dissipative and dispersive signal may therefore be expressed by
\begin{equation}\label{GammaEq}
\Gamma = \left|\frac{g_\text{e}\frac{\dif T}{\dif\gamma_\text{e}}\big|_\text{max}}{g_\text{om}\frac{\dif T}{\dif\omega_\text{o}}\big|_\text{max}}\right| = \frac{8 g_\text{e}(h) T_\text{d}(h)}{g_\text{om}(h) (1-T_\text{d}(h))},
\end{equation}
where $\Gamma > 1$ ($\Gamma < 1$) corresponds to a predominantly dissipative (dispersive) contribution to the $S^{1/2}_\text{VV}(\lambda)$ lineshape. This expression is plotted in Fig.\ \ref{fig:simulations}(b), using the simulated values for $g_\text{e}(h)$ and $g_\text{om}(h)$ (Fig.\ \ref{fig:simulations}(a)), for given $T_d$.  Note that in the limit of weak coupling ($1 - T_d \ll 1$), $\Gamma \gg 1$ is possible even when  $g_\text{e} < g_\text{om}$. Experimentally observed $T_\text{d}(h)$ for the device under study approximately follows the red line in Fig.\ \ref{fig:simulations}(b), which crosses from the dispersive to the dissipative regime ($\Gamma=1$) as $h$ is increased above $\sim$275 nm. This is consistent with the experimental observations in Fig.\ \ref{fig:optomechanics}(b) and (c), in which measurements show the prominence of dispersive coupling when the fiber is touching and dissipative when hovering.  Note that the respective values of $\Gamma$ for the hovering configurations (upper plots in Fig.\ \ref{fig:optomechanics}(c)) indicate that $h$ was larger for the M1 measurement than for M2:  when the taper is not stabilized through contact with the device, a slow drift in $h$ can occur despite the stage positions being fixed.

At first glance, the dramatic increase in $\Gamma$ with $h$ seems promising from an optomechanical cooling standpoint, especially for systems in the sideband-unresolved regime \cite{ref:elste2009qni}. Discussions of optomechanical cooling in systems exhibiting both dispersive and dissipative coupling \cite{ref:elste2009qni,ref:weiss2013qll} are typically couched in terms of the normalized coupling coefficients $\tilde{A} = \frac{-1}{\gamma_t} \frac{\dif\omega_o}{\dif x}x_\text{zpf}$ and 
$\tilde{B} = \frac{1}{\gamma_t}\frac{\dif\gamma_e}{\dif x}x_\text{zpf}$, where $x_\text{zpf}$ is the amplitude of the zero-point fluctuations of the mechanical oscillator.  The ratio $\tilde{B}/\tilde{A}=-g_\text{e}/g_\text{om}$ determines the optimal detuning required to minimize phonon occupation.  In contrast to $\Gamma$, $\tilde{B}/\tilde{A}<1$ for the entire $h$ range considered here (inset to Fig.\ \ref{fig:simulations}(b)), limiting the effectiveness of dissipative cooling. Further enhancement of the dissipative coupling strength, for example through more efficient fiber--nanocavity coupling \cite{ref:barclay2004eio}, is necessary for fiber-mediated tuning of the optomechanical coupling behavior described in this paper to effect dissipative cooling of the C modes of this device.

In summary, we have demonstrated how an optical fiber taper placed in the near field of a split-beam photonic crystal nanocavity can be used to renormalize the cavity mode, breaking its vertical symmetry and introducing additional optical decay channels.  This phenomenon enables readout of mechanical cantilever modes which otherwise have zero optomechanical coupling, and provides a spatially resolved probe of the device's mechanical modes, and the ability to tune the ratio of dissipative-to-dispersive coupling behavior.  Taken together, these effects have the potential to extend the range of device geometries used for optomechanics-based sensing applications.



\pagebreak
\widetext
\begin{center}
\textbf{\large Supplementary Material for ``Near-field tuning of optomechanical coupling in a split-beam nanocavity''}
\end{center}
\setcounter{equation}{0}
\setcounter{figure}{0}
\setcounter{table}{0}
\setcounter{page}{1}
\makeatletter
\renewcommand{\theequation}{S\arabic{equation}}
\renewcommand{\thefigure}{S\arabic{figure}}
\renewcommand{\bibnumfmt}[1]{[S#1]}
\renewcommand{\citenumfont}[1]{S#1}

\section{1. Dispersive and dissipative optomechanical coupling} 

In the following, we present a simplified model for fiber--cavity coupling in the sideband-unresolved regime, taking into account both dissipative and dispersive optomechanical coupling; a more complete discussion may be found in Appendix 1 of Ref.\ \cite{Sref:wu2014ddo}.

We consider an optical fiber taper placed in the near field of an optical cavity with resonance frequency $\omega_\text{o}$.  Input light from the fiber couples to the cavity, with the transmitted signal in the fundamental fiber mode being measured by a photodetector.  The loss channels for the cavity are:  1) coupling into the forward- or backward-propagating modes of the fiber, each described by a rate of $\gamma_\text{e}$, and 2)  intrinsic cavity loss ($\gamma_\text{i}$) and fiber-induced parasitic loss ($\gamma_\text{p}$), which we describe collectively by a rate of $\gamma_\text{i+p} = \gamma_\text{i} + \gamma_\text{p}$.

For weak fiber--cavity coupling, in which $\gamma_\text{e} \ll \gamma_\text{i+p}$, the transmission spectrum of the fiber may be written as \cite{Sref:wu2014ddo}
\begin{equation}
T \sim  
\frac{\Delta^2 + \left(\frac{\gamma_\text{i+p}}{2}\right)^2}{\Delta^2 + \left(\frac{\gamma_t}{2}\right)^2},
\label{eq:T}
\end{equation}
where $\Delta = \omega_\text{l} - \omega_\text{o}$ is the detuning of the input laser with respect to the cavity resonance, $\gamma_t = \gamma_\text{i+p} + 2\gamma_\text{e}$ is the total cavity optical loss rate, and we have neglected a Fano modification to the cavity response brought about by coupling to higher-order fiber modes that are converted to the fundamental fiber mode.
 
The cavity's optomechanical response with respect to a supported mechanical mode of frequency $\omega_\text{m}$ may be thought of as stemming from the effect of the global amplitude of the motion, $x$, on the parameters in Eq.\ (\ref{eq:T}).  In the sideband-unresolved/``bad-cavity'' regime ($\omega_\text{m} \ll \gamma_\text{t}$), the fiber transmission adiabatically follows the mechanical oscillation \cite{Sref:krause2012ahm}, such that a mechanical displacement $\dif x$ yields a corresponding change in transmission given by
\begin{equation}
\frac{dT}{dx}(\Delta) = \left | g_{\text{om}} \frac{\partial T}{\partial \Delta} + g_\text{i} \frac{\partial T}{\partial \gamma_\text{i+p}} + g_\text{e} \frac{\partial T}{\partial \gamma_\text{e}} \right |.
\label{eq:dTdx}
\end{equation}
In the above, $g_\text{om} = \dif\omega_\text{o}/\dif x$, $g_\text{i} = \dif\gamma_\text{i+p}/\dif x$, and $g_\text{e} =  \dif\gamma_\text{e}/\dif x$ are the dispersive, intrinsic dissipative, and external dissipative optomechanical coupling coefficients, respectively \cite{Sref:wu2014ddo}. 

Differentiating Eq.\ (\ref{eq:T}) with respect to each of its parameters yields
\begin{align}
 \frac{\partial T}{\partial \Delta} &= \frac{2\Delta(1-T)}{\Delta^2 + \left(\gamma_t/2\right)^2}\\
	\frac{\partial T}{\partial \gamma_\text{i+p}} &= \frac{\gamma_\text{i+p} - T (\gamma_\text{i+p} + 2\gamma_\text{e})}{\Delta^2 + \left(\gamma_t/2\right)^2}\\
 \frac{\partial T}{\partial \gamma_\text{e}} &= \frac{ - 2\gamma_t T}{\Delta^2 + \left(\gamma_t/2\right)^2}.
\label{eq:dTdgi}
\end{align}
We may then compare the contribution of these terms to the $S_\text{VV}(\lambda)$ lineshape by considering their peak amplitudes \cite{Sref:krause2012ahm}: 
\begin{align}
\frac{\partial T}{\partial \Delta} \bigg|_\text{max} &= \frac{\dif T}{\dif\Delta}(\Delta = \frac{\gamma_t}{2}) = (1-T_{\text{d}}) \frac{Q_\text{o}}{\omega_{o}}\\
\frac{\partial T}{\partial \gamma_\text{i+p}} \bigg|_\text{max} &= \frac{\dif T}{\dif\gamma_\text{i+p}}(\Delta = 0) = 4 (1-T_{\text{d}}) \frac{Q_\text{o}}{\omega_{o}}\\
\frac{\partial T}{\partial \gamma_\text{e}} \bigg|_\text{max} &= \frac{\dif T}{\dif\gamma_\text{e}}(\Delta = 0) = -8 T_{\text{d}} \frac{Q_\text{o}}{\omega_{o}}
\end{align}
where $T_\text{d} = \gamma_\text{i+p}^2/\gamma_\text{t}^2$ is the on-resonance transmission depth and $Q_\text{o} = \omega_\text{o}/\gamma_\text{t}$ is the optical quality factor.  For weak fiber--nanocavity coupling ($1 - T_\text{d} \ll 1$) \footnote{Note that in this regime contributions to the optomechanical signal from $\dif\gamma_\text{p}(x)/\dif x$ will be small compared to $\dif\gamma_\text{e}(x)/\dif x$ if $|\dif\gamma_\text{p}/\dif x| \le | \dif\gamma_\text{e}/\dif x|$.}, and for the mechanical modes with minimal intrinsic dissipative optomechanical coupling ($\dif\gamma_\text{i}/\dif x \ll \dif\gamma_\text{e}/\dif x$), as in the system studied here, the dissipative contribution to the measured optomechanical signal is dominated by $\gamma_\text{e}$ \cite{Sref:wu2014ddo}. As such, we express the relative balance of experimentally observed dissipative and dispersive signal by the ratio $\Gamma$ given in the main text:
\begin{equation}\label{GammaEq}
\Gamma = \left|\frac{g_\text{e}\frac{\dif T}{\dif\gamma_\text{e}}\big|_\text{max}}{g_\text{om}\frac{\dif T}{\dif\omega_\text{o}}\big|_\text{max}}\right| = \frac{8 g_\text{e} T_\text{d}}{g_\text{om} (1-T_\text{d})}.
\end{equation}


\section{2. FEA estimate of $g_\text{e}$} 

The system under study is a fiber--nanocavity system in which the distance $h$ changes the coupling rate $\gamma_\text{e}(h)$ between the cavity and both the forward- and backward-propagating waves of the fiber as illustrated in Fig. 1(b) of the main text. Here $h$ is defined as the distance between the nanocavity and the outer boundary of the fiber. The presence of the fiber may also create other loss channels by scattering light away from the fiber or by coupling to higher-order waveguide modes.  These are bundled together as parasitic loss rate $\gamma_\text{p}(h)$ \cite{Sref:spillane2003ift}. The nanocavity itself has a radiation loss rate of $\gamma_\text{rad}$, which can be computed via numerical simulations (FDTD, FEA), and a scattering loss rate $\gamma_\text{s}$ due to fabrication imperfections.  Together, they form the intrinsic loss of the nanocavity:   $\gamma_\text{i} = \gamma_\text{rad} + \gamma_\text{s}$.  The total loss rate $\gamma_\text{t}$ is then given by
\begin{equation}
	\gamma_\text{t}(h) = \gamma_\text{i} + \gamma_\text{p}(h) + 2\gamma_\text{e}(h).
\end{equation}

To generate the theoretical values for $g_\text{e}(h)$, and hence $T_\text{d}(h)$ and $\Gamma(T_\text{d},g_\text{e},g_\text{om})$, used in Fig.\ 4 of the main text,  $\gamma_\text{e}(h)$ was estimated from FEA (COMSOL) simulations of $\gamma_\text{t}(h)$.  Precisely determining $\gamma_\text{e}(h)$ given $\gamma_\text{t}(h)$ requires knowledge of $\gamma_\text{p}(h)$.  Here, we assess $\gamma_\text{p}$ based on experimentally observed $T_\text{d}(h=0)$ and $\gamma_\text{i}=\gamma_\text{t}(h\to\infty)$, from which the ratio  $\gamma_\text{e}(0)/(\gamma_\text{e}(0)+\gamma_\text{p}(0)) = 0.4$ was extracted.  Making the simplifying assumption that $\gamma_\text{e}/(\gamma_\text{e}+\gamma_\text{p})$ is constant for all $h$ allows an estimate of $\gamma_\text{e}(h)$ to be determined from the simulated values of $\gamma_\text{t}(h)$.  

This procedure likely overestimates $\gamma_\text{p}$ for $h > 0$, as $\gamma_\text{p}$ typically decays with $h$ quickly compared to $\gamma_\text{e}$, i.e.\ the coupling becomes more ideal as $h$ increases \cite{Sref:spillane2003ift}.  As a result, this procedure may underestimate $\gamma_\text{e}$ for large $h$, and underestimate the decay constant $\Lambda_\text{e}$ of $g_\text{e}$. However, the key features in Figure 4, notably that $\Lambda_\text{e} \gg \Lambda_\text{om}$, and that $\Gamma = 1$ when $h\sim 275$ nm, are not found to be significantly affected by these uncertainties. In principle, $\gamma_\text{p}(h)$ may be measured experimentally; however, this was difficult in the system under study due to the relatively small $h < 500$ nm at which coupling was observed (resulting in significant fiber taper insertion loss) and the poor contrast of the measured nanocavity resonance. In future, fabrication of nanocavities with higher $Q_\text{o}$ may address this difficulty.

\section{3. Perturbative approximations for $g_\text{om}$ and $g_\text{e}$}

To gain insight into the physical mechanisms governing the effect of the fiber taper on $g_\text{om}$ and $g_\text{e}$, we evaluate the shift in cavity resonance frequency, $\omega_o$, and coupling rate between the fiber and cavity, $\gamma_e$, using first-order perturbation theory.

\subsubsection{A.  Unperturbed cavity and fiber fields}

\begin{figure}[h] 
\begin{center}
\epsfig{figure=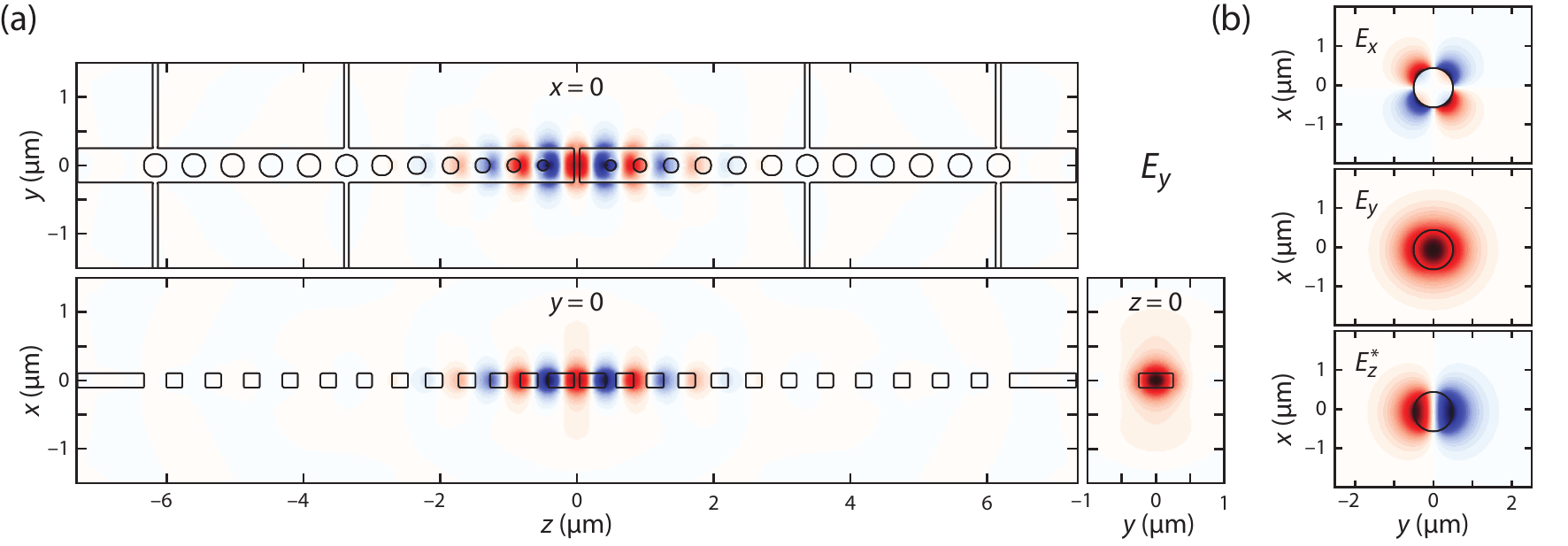, width=18cm}
\caption{Unperturbed electric field profiles of (a) split-beam cavity ($E_y$ only) and (b) 1-$\mu$m-diameter fiber}
\label{fig:fields}
\end{center}
\end{figure}

The unperturbed cavity field $\mathbf{E}_\text{c}$, the dominant $y$-component of which is shown in Fig.\ \ref{fig:fields}(a), was calculated using FDTD simulations \cite{Sref:oskooi2010mff} of the cavity geometry as determined from SEM images of the device.  The dielectric profile of the cavity, $\epsilon_\text{c}(\mathbf{r})$, is assumed to have inversion symmetry; in particular, the circular hole radii and positions are specified to be symmetric with respect to the $z=0$ plane.  The fundamental TE-like cavity mode ($E$-field even in $x$, odd in $y$) has a resonance wavelength of $\sim$1612 nm, a quality factor of $1.2\times10^4$ (limited by scattering in the $x$ and $y$ directions), and an effective mode volume of $\sim0.35\;(\lambda/n)^3$. 

The unperturbed fiber taper fields were calculated using a frequency-domain eigenmode solver \cite{Sref:johnson2001bif}, assuming a SiO$_2$ ($n_\text{f} = 1.44$) fiber with a diameter of 1 $\mu$m in air.  This fiber supports a single TE-like mode at a wavelength of 1612 nm, with a propagation constant $\beta = 4.5~\mu \text{m}^{-1}$.  

\subsubsection{B.  Cavity resonant frequency}

\begin{figure}[h] 
\begin{center}
\epsfig{figure=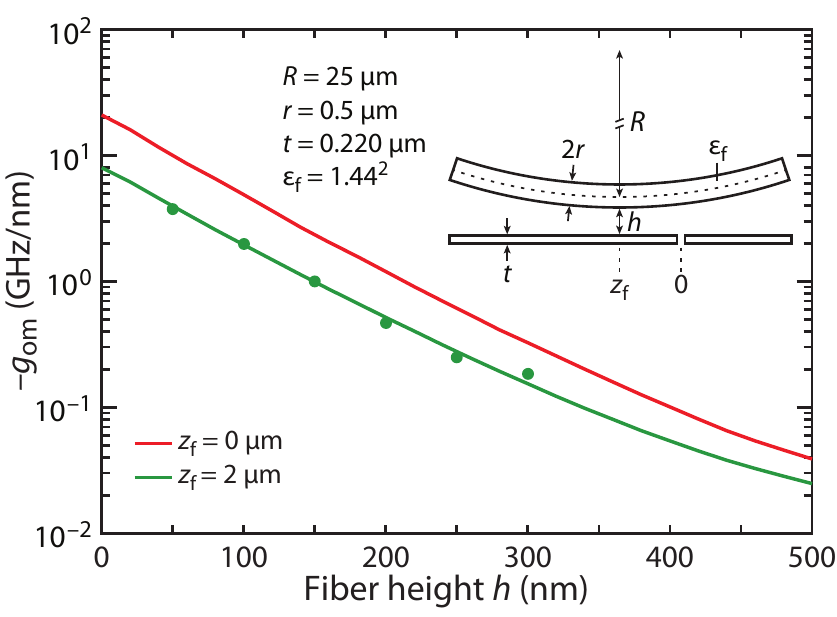, width=8.5cm}
\caption{Dispersive coupling coefficient $g_\text{om}$ calculated using first-order perturbation theory; the simulation geometry is shown in the inset.  The red (green) line corresponds to an axial fiber offset $z_\text{f}$ of 0 ($-2$) $\mu$m from the cavity center.   The green circles are the $g_\text{om,M1}$ values from Fig.\ 4(a) of the main text, calculated using FEA.}
\label{fig:gom_pert}
\end{center}
\end{figure}

The first-order correction to the resonant frequency of an electromagnetic cavity due to a change in permittivity may be calculated using \cite{Sref:johnson2002ptm}
\begin{equation}\label{pertw}
\Delta\omega_o=-\frac{\omega_o}{2}\frac{\langle\mathbf{E}_\text{c}|\Delta\epsilon_\text{f}|\mathbf{E}_\text{c}\rangle}{\langle\mathbf{E}_\text{c}|\epsilon_\text{c}|\mathbf{E}_\text{c}\rangle},
\end{equation}
where $\mathbf{E}_\text{c}$ and $\omega_o$ are the unperturbed cavity electric field and resonant frequency, respectively, $\Delta\epsilon_\text{f}$ is the perturbation of the local dielectric environment due to the fiber, and $\langle\,\rangle$ represents integration over all space.  For the geometry considered in this paper, $\Delta\epsilon_\text{f} = \epsilon_\text{f} - 1$, with the integral restricted to the region inside the fiber taper.  As in the finite-element calculations in Fig.\  4(a) of the main text, we model a dimpled fiber with a 25 $\mu$m radius of curvature (see inset to Fig.\ \ref{fig:gom_pert}).

From this expression, we see that the change in the cavity resonant frequency with $h$ scales with the intensity of the evanescent cavity field overlapping with the fiber.  For a cantilever mode, $\dif x \equiv -\dif h$, such that $g_\text{om} \sim -\frac{\dif\Delta\omega_\text{o}}{\dif h}$ decays with the same quasi-exponential dependence.  Fig.\ \ref{fig:gom_pert} plots $g_\text{om}$ using this approach for the dimple centered on the cavity ($z_\text{f}=0~\mu$m) and offset axially over one of the mirrors  ($z_\text{f}=-2~\mu$m); the latter agrees well with $g_\text{om}$ calculated using FEA for the full fiber--cavity system, as shown in Fig.\ 4(a) of the main text.

\subsubsection{C.  Fiber--cavity coupling}

\begin{figure}[h] 
\begin{center}
\epsfig{figure=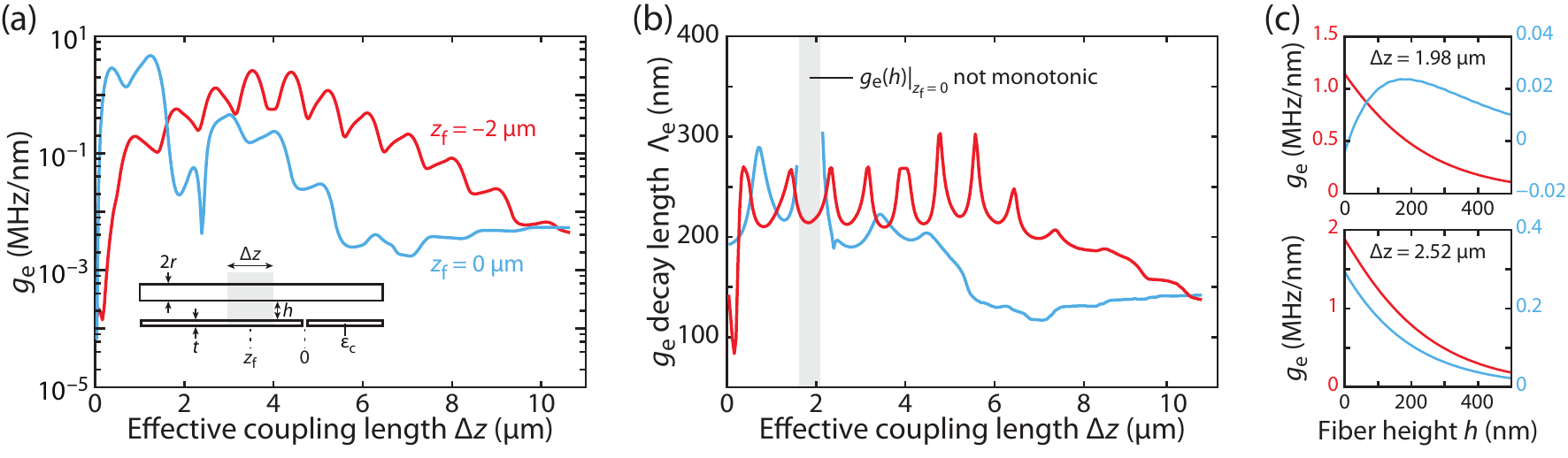, width=1\linewidth}
\caption{Dissipative coupling coefficient $g_\text{e}$ calculated from mode-coupling theory for a fiber height of 200 nm; the simulation geometry is shown in the inset to (a), where the grey box denotes the integration limits on $z$.  In all cases, blue (red) corresponds to a $z_\text{f}$ of 0 ($-2$) $\mu$m. (a) $g_\text{e}$ as a function of effective coupling length. (b) Decay lengths obtained from single-exponential fits of $g_\text{e}(h)$. (c) $g_\text{e}$ vs.\ $h$ for coupling lengths of 1.98 $\mu$m (top) and 2.52 $\mu$m (bottom).}
\label{fig:ge_pert}
\end{center}
\end{figure}

An approximation for the cavity loss rate into the fiber, $\gamma_\text{e}$, can be obtained from coupling-mode analysis for a generalized waveguide--resonator system \cite{Sref:manolatou1999cma}.  Neglecting dispersion, the loss rate into either the forward- or backward-propagating fiber mode is
\begin{equation}\label{kappa}
\gamma_e=\left|\frac{\omega\epsilon_0}{4}\int_{z_1}^{z_2}\!\dif z\iint\dif x\;\dif y\;(\epsilon_\text{c}-1)\;\mathbf{E}_\text{c}^*\cdot\mathbf{E}_\text{f}\;e^{-i\beta z}\right|^2,
\end{equation}
where $\epsilon_\text{c}$ is the relative permittivity of the cavity, $\mathbf{E}_\text{c}(x,y,z)$ is the unperturbed cavity electric field distribution (normalized to unit energy), $\mathbf{E}_\text{f}(x,y)$ is the unperturbed fiber electric field mode profile (normalized to unit power), $\beta$ is the fiber mode propagation constant, and the integrals in $x$ and $y$ are restricted to the region inside the cavity dielectric.  As a simple approximation for  the effect of the dimple curvature, we assume a straight fiber at a distance $h$ above the cavity and integrate over an effective coupling length $\Delta z$ centered at $z_\text{f}$ (i.e,. $z_1=z_\text{f}-\frac{\Delta z}{2}$,  $z_2=z_\text{f}+\frac{\Delta z}{2}$).  Assuming $\dif x>0$ corresponds to deflection of the cantilever toward the fiber, then for a fiber--cavity separation $h$, we then have $g_\text{e}\sim-\frac{\dif\gamma_e}{\dif h}$.  Fig.\ \ref{fig:ge_pert}(a) plots $g_\text{e}$ calculated via this approach for dimple center positions $z_\text{f}$ of 0 and $-2$ $\mu$m.

Note that $g_\text{e}$ calculated using this approach does not take into account contributions from coupling to higher-order fiber modes that are converted to the fundamental mode, which in explains its lower magnitude with respect to the calculation shown in Fig.\ 4(a) of the main text, and as explained in \S2 above.  Although this treatment is approximate, it captures several features of the full FEA approach, including non-monotonic behavior of $g_\text{e}(h)$ with $z_\text{f}=0~\mu$m for certain coupling lengths (Fig.\ \ref{fig:ge_pert}(c), top), and sensitivity of the magnitude of $g_\text{e}$  (Fig.\  \ref{fig:ge_pert}(a)) and its decay length $\Lambda_\text{e}$  (Fig.\ \ref{fig:ge_pert}(b)) to dimple position $z_\text{f}$.   The richer physics of this coupling mechanism compared with $g_\text{om}$ may be traced to its origin as an interference effect, which does not enter into the dispersive coupling calculation.


\end{document}